\documentclass{pasj00}
\draft
\nonstopmode

\SetRunningHead{Y. Matsumoto et al.}{Gamma-Ray Emitting Solar Flares}

\begin{document}

\title{A Statistical Study of Gamma-Ray Emitting \\ Solar Flares Observed with Yohkoh}

\author{
Yukari {\sc Matsumoto},$^1$ Kazuo {\sc Makishima},$^{1,2}$ 
Jun'ichi {\sc Kotoku},$^{1,2}$ Masato {\sc Yoshimori},$^3$\\
Kazuharu {\sc Suga},$^3$ Takeo {\sc Kosugi},$^4$ 
Satoshi {\sc Masuda}$^5$, and Kouji {\sc Morimoto}$^6$ \\
{\small {\it 1 Department of Physics, The University of Tokyo, 
                7-3-1 Hongo,  Bunkyo-ku,  Tokyo 113--0033}}\\
\centerline{\small {maxima@phys.s.u-tokyo.ac.jp}}\\
{\small {\it 2 Cosmic Radiation Laboratory, The Institute of Physical and Chemical Research (RIKEN),}}\\
{\small {\it  2-1 Hirosawa, Wako, Saitama 351-0198}}\\
{\small {\it 3 Department of Physics, Rikkyo University, 3-34-1 Nishi-Ikebukuro, Toshima-Ku,
                Tokyo 171-8501}}\\
{\small {\it 4 Institute of Space and Astronautical Science, 
                3-1-1 Yoshinodai, Sagamihara, Kanagawa 229--8510}}\\ 
{\small {\it 5 Solar-Terrestrial Environment Laboratory, Nagoya University, 
                Toyokawa, Aichi 442-8507}}\\
{\small {\it 6 RI Beam Science Laboratory, 
The Institute of Physical and Chemical Research Research (RIKEN),}}\\
{\small {\it  2-1 Hirosawa, Wako, Saitama 351-0198}}\\
}
\Received{2002 December 25}
\Accepted{2004 November 17}
\maketitle

\begin{abstract}
Gamma-ray emitting solar flares observed with Yohkoh 
were analyzed from a statistical viewpoint.
The four-band hard X-ray (15--95 keV) photometric data,
taken with the Hard X-ray Telescope onboard Yohkoh,
were utilized in combination with the 
spectro-photometric gamma-ray (0.2--30 MeV) data
obtained with the Gamma-Ray Spectrometer.
The GOES class was also incorporated.
Out of 2788 X-ray flares observed from 1991 October to 2001 December,
178 events with strong hard X-ray emission were selected.
Among them, 40 flares were further
found to show significant gamma-ray emission.
A fractal dimension analysis and multi-band 
color--color plots of the 40 flares suggest
that their soft X-ray to MeV gamma-ray spectral energy 
distributions involve at least four independent parameters.
These are: 
(1) the overall flare size; 
(2) the relative intensities of the thermal vs. non-thermal signals;
(3) the gamma-ray to hard X-ray intensity ratio;
and
(4) the hard X-ray spectral slope.
These results are examined for possible selection effects.
Also, the meanings of the third parameter are briefly considered.
\end{abstract}

\thispagestyle{headings}
\section{Introduction}

A decade of solar observations with Yohkoh 
(Ogawara et al. 1991; Acton et al. 1992) 
has greatly reinforced the view
that solar flares are powered by magnetic reconnection,
which takes place in the X-shaped field region 
located above a magnetic loop system 
(e.g., Carmichael 1962; Sturrock, Coppi 1966; 
Hirayama 1974; Kopp, Pneuman 1976).
The reconnection produces an outward mass ejection,
as well as a downward plasma flow,
which streams down along the magnetic field lines.

Energetic electrons in such a down-stream emit non-thermal bremsstrahlung,
mainly from loop footpoints.
Since the bremsstrahlung continuum extends well into gamma-ray regions,
the electrons must acquire a non-thermal energy distribution
extending up to MeV (or sometimes GeV) energies, 
before they impinge onto the chromosphere.
Nevertheless, we do not yet know
where they acquire such high energies.
The non-thermal electron energy distribution 
may have already been produced at the reconnection site.
Alternatively, the down-stream may be initially cool,
and randomized via a standing shock (or similar processes) 
immediately above the chromosphere. 
Or else, such a secondary acceleration may take place elsewhere 
in between the reconnection point and the main emission region
at the loop footpoints.
Even a single flare may well involve these multiple sites 
with different acceleration mechanisms.

In this regard, important new results have been achieved with 
the Yohkoh HXT (Hard X-ray Telescope; Kosugi et al. 1991, 1992),
which can take flare images in the 14--93 keV hard X-ray range
with an unprecedented angular resolution of $\sim 5''$.
The HXT has in particular allowed for the detection of 
hard X-ray sources located at the top of magnetic loop systems 
(Masuda 1994; Masuda et al. 1994, 1995; Petrosian et al. 2002).
This suggests an intriguing possibility
that particles are accelerated 
as the downward plasma flow collides with the magnetic loops,
although we do not yet know whether the loop-top hard X-ray sources
are of non-thermal origin or not.

In order to better understand the physics of 
particle acceleration in solar flares,
we are systematically analyzing the HXT data,
for the first time jointly with those from 
the Yohkoh GRS (Gamma-Ray Spectrometer; Yoshimori et al. 1991),
which performs spectro-photometry over the 0.2--30 MeV band.
In the present paper, 
which is meant to be the first of a series,
we describe a statistical study of the entire sample of 
gamma-ray emitting solar flares observed with Yohkoh.
Section 2 is devoted to a brief description of the HXT and GRS instrumentations.
The sample consisting of 40 flares is defined in section 3.
In section 4, we analyze the 40 flares for
their broad-band spectral energy distributions (SEDs), 
over the soft X-ray to MeV gamma-ray energies.
The results are discussed in section 5.

\section{Instrumentation}

\subsection{The HXT (Hard X-ray Telescope)}

The HXT (Kosugi et al. 1991) is a hard X-ray imager,
incorporating 64 ``modulation subcollimators'' 
in the Fourier-synthesis configuration.
It can take hard X-ray images of solar flares every 0.5 s,
simultaneously in four energy bands:
L-band (14--23 keV), M1-band (23--33 keV), M2-band (33--53 keV) and H-band (53--93 keV). 
The angular resolution attains an unprecedented level of $\sim 5''$,
on the condition that the statistics allows and the source size does not exceed $\sim 2'$.

The HXT image synthesis utilizes the maximum-entropy method
(e.g., Kosugi et al. 1991, 1992; Sakao et al. 1992).
This requires a highly accurate (typically to 1\%) equalization
of boundaries of the four energy channels among the 64 subcollimators.
Otherwise, they would sample different energy intervals 
of steeply falling flare spectra,
and hence their counts would contradict one another.
These adjustments have occasionally been carried out in orbit
by trimming, via commands, 
the relative gains of the 64 detectors
in reference to built-in $^{241}$Am calibration isotopes.

The HXT image synthesis also requires accurate 
(typically $\sim 1''$ in terms of the angular response) knowledge 
of the modulation patterns of the 64 subcollimators. 
This was initially done by Sakao (1994), Inda-Koide (1994), 
and Inda-Koide et al. (1995), based mainly on pre-launch measurements.
Then, Sato (1997) and Sato et al. (1999) ``self-calibrated'' 
the subcollimator parameters using the solar flares themselves,
and significantly improved the quality of the synthesized images.
As a result of these efforts, 
the HXT not only provides high-resolution images of solar flares,
but also allows us to study spatially resolved spectra of the flare hard X-rays. 

In the present paper, however, we utilize the HXT 
only as a four-band fast photometer with  accurate energy calibration,
rather than an imaging spectrometer.
This can be done simply by summing up 
the output counts from the 64 subcollimators.
Because each subcollimator is coupled to 
a relatively small (1 inch square) Na I scintillator 
and an independent fast electronics chain,
the HXT has a very wide (exceeding 3 orders of magnitude) photometric dynamic range,
being free from signal pile-up even for very large flares.
We will fully utilize the HXT images in the subsequent publications.

\subsection{The GRS (Gamma-Ray Spectrometer)}

The GRS (Yoshimori et al. 1991), 
which constitutes the Wide-Band Spectrometer system of Yohkoh,
is a gamma-ray spectro-photometer.
It consists of two identical BGO scintillation detectors,
named GRS-1 and GRS-2,
each having a 7.6 cm diameter and a 5.1 cm thickness.
In the low-energy range (typically below 10 MeV),
each of GRS-1 and GRS-2 provides 128-channel 
pulse-height data (PH-L) every 4 s, 
and 4-channel broad-band pulse-count data (PC1 through PC4) every 0.25 or 0.5 s.
Above $\sim 10$ MeV, 
each detector provides 16-channel pulse-height data (PH-H) every 4 s, 
and 2-channel pulse-count data (PC5 and PC6) every 0.5 s. 

The GRS detector gain somewhat changed in orbit, 
over a year after the launch.
We therefore re-analyzed all of the in-flight calibration data,
and updated the detector response (Matsumoto 2002).
Table~\ref{tbl:GRS} summarizes the revised energy ranges of the GRS PC data,
which are valid after 1997.

Although the GRS has been calibrated to a fair precision
and the HXT to a still higher accuracy,
the energy rages of these two instruments do not overlap,
and there is no canonical flare spectra available for  their mutual calibration.
As a result, their relative sensitivities for typical solar flares
are considered to be uncertain up to a factor of $\sim 2$.

\section{The Sample Definition}

The Yohkoh HXT flare list ver.2001-07 
(Sawa et al. 2001, National Astronomical Observatory, Japan)
compiles background-subtracted peak counts in the four HXT energy bands,
for all 2359 flares observed by Yohkoh from 1991 October to 2001 June.
This period  covers the second half of solar cycle 21 
and the first half of solar cycle 22.
By 2001 December when the mission came to an end,
Yohkoh further detected 429 more flares,
thus making the total flare number 2788.

Among the 2788 flares, 
217 events exhibit hard X-ray emission strong enough for image synthesis,
exceeding 100 cts s$^{-1}$ SC$^{-1}$ in at least one of the four HXT bands,
where SC stands for subcollimator.
The flare number reduces to 178
when we exclude those events 
of which the beginning was not recorded by Yohkoh,
and those which occurred when either the HXT or the GRS was not operational.
These 178 flares constitute our ``Preliminary Sample'' (Matsumoto 2002). 

We then examined our Preliminary Sample for gamma-ray signals, 
and found 40 events of which the GRS-(1+2) PC1 peak count rate,
after background subtraction, 
was higher than 100 cts s$^{-1}$,
which typically corresponded to 4$\sigma$ of the photon counting errors in 2 seconds.
These 40 gamma-ray emitting flares define our ``Main Sample'';
table \ref{tbl:sample} summarizes them,
in terms of their peak counts in each of the HXT and GRS energy bands.
Here, the gamma-ray intensities refer to the GRS-1 plus GRS-2 counts,
although their energy ranges somewhat differ from each other (table~\ref{tbl:GRS}).
Note that the flare peaks in different energy bands
are not necessarily simultaneous.

\section{Broad-Band Spectral Energy Distributions of the Sample Flares}

Here, we describe some statistical studies of the 40 
gamma-ray flares in our Main Sample (table~\ref{tbl:sample}),
defined in the previous section.
In particular, we examine
whether their broadband SEDs are characterized by a small number of parameters
with clear physical meanings.

\subsection{Correlation among Different Energy Bands}

\begin{figure}[phtpb]
\centerline{\includegraphics[width=16cm,angle=0,clip]
{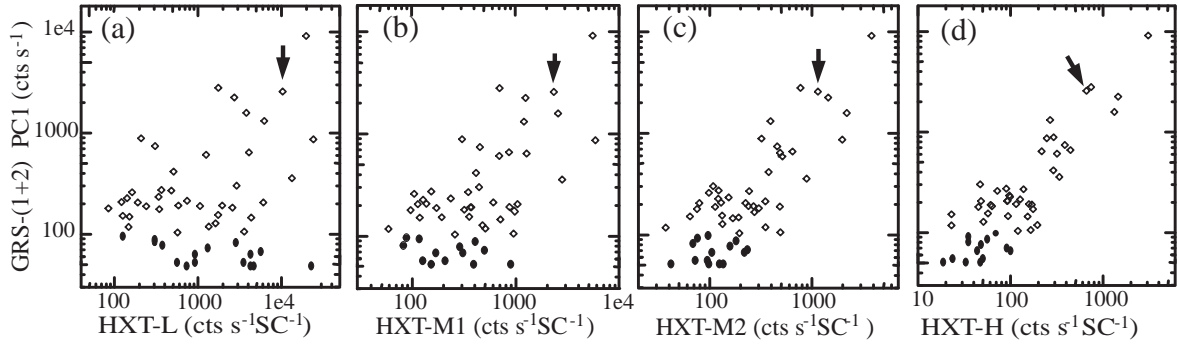}}
\caption{
Correlations between the GRS-(1+2) PC1 peak counts
and those in several other energy bands,
plotted for the gamma-ray emitting flares observed with Yohkoh.
In each figure, one data point represents one solar flare.
Open symbols represent the 40 gamma-ray flares 
of table~\ref{tbl:sample} (``Main Sample''),
while filled symbols those in ``Additional Sample''
with the peak GRS-(1+2) PC1 counts in the range  50--99 cts s$^{-1}$.
The arrows indicate the 1998 August 18 flare.
}
\label{fig:scatter}
\end{figure}

Over the 40 gamma-ray flares,
we calculated correlation coefficients between 
the peak intensities in different energy bands.
We utilized various combinations among 7 of the 
8 spectral bands given in table~\ref{tbl:sample}: 
GOES class, the four HXT bands, and the PC1 and PC2 bands of GRS-(1+2).
We did not use PC3 or higher-energy bands, 
because of the poor statistics.
The obtained correlation coefficients are presented in table~\ref{tbl:correlation_coef}.
In figure~\ref{fig:scatter}, 
we show some scatter plots of the GRS PC1 peak counts
against those in other spectral channels.

As can be seen in table~\ref{tbl:correlation_coef} and figure~\ref{fig:scatter},
the peak counts between two wide-apart energy bands are poorly correlated,
whereas the correlation increases 
as the two bands become closer along the energy axis.
In particular, the flare gamma-ray intensity 
is relatively well-correlated with the HXT-M2 and HXT-H peak counts,
implying that the gamma-ray emission is generally accompanied by hard X-ray emission. 

Among these flares, 
a particularly interesting event is the GOES-class X4.9 flare,
which occurred at the east solar limb (N30E84) 
on 1998 August 18 22:14 UT (Petrosian et al. 2002).
In the GRS-PC3 and higher energy ranges,
it is by far the strongest flare that Yohkoh has ever detected.
Since this particular event plays an important role in our subsequent papers,
we indicate it in figure~\ref{fig:scatter} with arrows.

\subsection{Fractal Dimension Analysis}
The large scatters between the flare intensities in  different energy bands
(table~\ref{tbl:correlation_coef}) imply
that flares can take a variety of broad-band SEDs.
Then, how many independent parameters control the distribution?
Obviously, the most dominant freedom is the overall flare size.
There can still be other parameters, 
such as the relative dominance between thermal and non-thermal signals, 
the spectral slope of the hard X-ray emission, and so on.

We may study this issue by so-called fractal dimension analysis.
We consider a 7-dimensional linear space,
spanned by the 7 energy bands utilized in table~\ref{tbl:correlation_coef}.
In this linear space, we express the $i$-th ($i=1,2,..,40$) flare with a vector,
\begin{equation}
  \vec{V}_i = \left( \frac{P_{i,1}}{A_1},  \frac{P_{i,2}}{A_2},...,
             \frac{P_{i,7}}{A_7} \right) .
\end{equation}
Here, $P_{i,k}$ is its peak intensity in the $k$-th energy band,
with $k=1$ the GOES class and $k=7$ the GRS-(1+2) PC2 counts,
while $A_k \equiv \Sigma^{40}_{i=1} P_{i,k} / 40$
is the flare-averaged peak count in the $k$-th band.
Thus, the seven variables are normalized for
each flare vector to have on average the same length along the seven axes,
but the overall vector length scatters greatly among the 40 flares,
reflecting the flare size.
To eliminate this dominant freedom,
we further normalize each $\vec{V}_i$ to have a length of $\sqrt7$,
and write the normalized vector as
\begin{equation}
  \vec{v}_i = \vec{V}_i \times \sqrt{
                \frac{7}{\Sigma^{7}_{k=1} ({P_{i,k}}/A_k)^2}
                                    }.
\end{equation}
In the 7-dimensional embedding space,
the set of 40 vectors, $\{ \vec{v}_i; i=1,2,..,40 \}$,
forms a ``subspace'' {\bf S} centered on 
the mean vector $\vec{\langle v \rangle} \sim (1,1,1,1,1,1,1)$.

If the SED of our sample flares has 
$n$ independent degrees of freedom,
we expect the subspace {\bf S} to have a dimension of $n-1$ ($n \le 7$),
since we have already suppressed one obvious degree of freedom, the flare size.
To actually calculate the dimension of {\bf S},
we define the ``distance'' of the $i$-th flare from the mean vector, as
\begin{equation}
r_i = \Sigma^{7}_{k=1} \left\{ \left(v_{i,k} - \langle v_k \rangle \right)/\sigma_k \right\}^2 .
\end{equation}
Here, $v_{i,k}$ and $\langle  v_k \rangle $ are 
the $k$-th component of $\vec{v_i}$ and $\vec{\langle v \rangle}$,
respectively, and $\sigma_k$ is the standard deviation of $v_{i,k}$
calculated over $i=1,2,..,40$.
A large distance indicates
that the spectral shape of this particular flare
is very much deviated from the mean SED.
Then, we define a cumulative flare number distribution, $C(<r)$,
as the number of flares, 
of which $r_i$ is smaller than a given value $r$.
We expect $C(<r) \propto r^{n-1}$,
because the ``hyper-volume'' of a subspace of dimension $n-1$
should scales as $\propto r^{n-1}$.

Figure \ref{fig:fractal}a shows $C(<r)$ as a function of the distance $r$,
calculated for the 40 gamma-ray flares.
Thus, the logarithmic slope of $C(<r)$ depends 
to some extent on the distance region to be employed.
If we discard the region of $r<1.8$ where the number statistics are poor
and the region of $r>3.0$ where the radius gets close to the distribution periphery,
the data are fitted by a logarithmic slope of $n-1 = 2.8 \pm 0.8$, 
implying $n \sim 4$.
That is, the SEDs (including its normalization) of the 40 flares
measured in the seven broad energy bands are
specified by approximately 4 independent parameters.

\subsection{Color--Color Diagrams}
\begin{figure}[phbt]
\centerline{\includegraphics[width=16.5cm,angle=0,clip]
{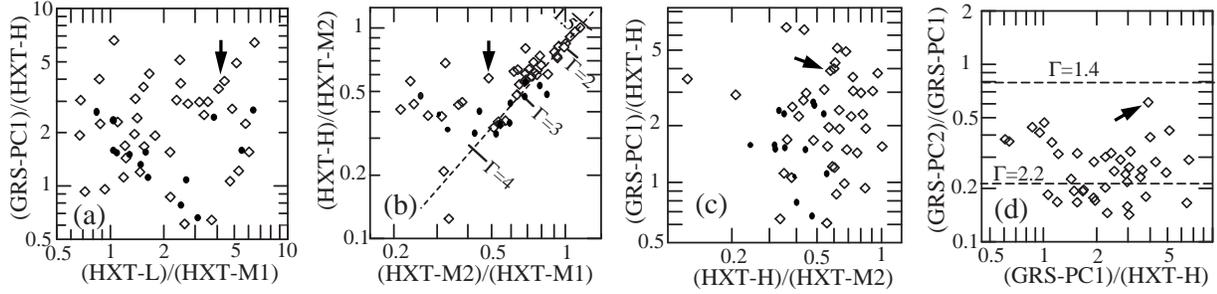}}
\caption{Several color--color plots of the gamma-ray emitting flares.
Meanings of the open and filled symbols are 
the same as in figure~\ref{fig:scatter}.
Arrows indicate the 1998 August 18 flare.
(a) Abscissa shows ratios of the HXT-L vs. HXT-M1 peak counts,
while ordinate shows that of the GRS-(1+2) PC1 vs. HXT-H peak counts.
(b) A similar plot, but between the HXT-M2/HXT-M1 ratio and the HXT-H/HXT-M2 ratio.
The dashed diagonal line indicates the locus of a single power-law spectrum,
with tick marks indicating a photon index of 1.5, 2.0, 3.0, and 4.0.
(c) Between the HXT-H/HXT-M2 ratio and the GRS-(1+2) PC1 vs. HXT-H ratio.
(d) Between the GRS-(1+2) PC1 vs. HXT-H ratio and the GRS-(1+2) PC2/PC1 ratio.
Two dashed horizontal lines indicate a power-law photon index of 1.4 and 2.2
for the GRS data.}
\label{fig:color_color}
\end{figure}

In order to consider the meanings of the four parameters
that have been found to control the broad-band SED of the flares in our Main Sample, 
we have arranged them in figure~\ref{fig:color_color}
in the form of several ``color--color'' plots.
There, the peak-count ratios between a particular pair of energy bands
are plotted against those between other band pairs.
By taking ratios, we can remove the obvious 1st degree of freedom, 
i.e., the overall flare size.

Figure~\ref{fig:color_color}a compares the GRS-PC1 vs. HXT-H peak-count ratios,
against those between HXT-L and HXT-M1.
Thus, the 40 flares exhibit a fully two-dimensional distribution
on this particular parameter space,
with nearly an order-of-magnitude scatter in both colors.
This immediately uncovers the presence of two more (2nd and 3rd) parameters.
The 2nd one is the strength of thermal X-ray signals relative to the hard X-ray signal,
as represented by the HXT-L/HXT-M1 ratio (abscissa of figure~\ref{fig:color_color}a).
The 3rd one is the gamma-ray intensity relative to the hard X-ray signal,
as represented by the GRS-PC1/HXT-H ratio (ordinate of fig~\ref{fig:color_color}a).

The 3rd parameter thus identified should be 
distinguished from the relatively tight correlation 
seen between the hard X-ray and gamma-ray counts 
on a larger scale (figure~\ref{fig:scatter}),
which is caused by the dominant 1st degree of freedom.
After removing the large-scale correlation, 
the gamma-ray to hard X-ray intensity ratio exhibits a residual scatter.
Namely, a flare with a strong hard X-ray signal 
also tends to exhibit a strong gamma-ray emission,
but the correlation is not tight enough to
exclude room for the 3rd parameter to operate.

When the 40 flares are plotted on the color--color plane
formed by the three consecutive hard X-ray bands (M1, M2, and H) of the HXT,
the distribution becomes significantly different;
as can be seen in figure~\ref{fig:color_color}b, 
a major fraction of events align almost one-dimensionally.
Furthermore, this alignment occurs along the dashed diagonal line,
which represents the locus of a family of model spectra,
each extending from 10 to 100 keV with a single power-law slope.
As to these flares, the M2/M1 and H/M2 ratios are therefore
specified simultaneously by a single parameter, 
i.e., the spectral slope of the hard X-ray emission.
The implied photon indices vary from 1.5 to 5,
with a considerable clustering in the range 2.5--3.0.
Although about one-third of the events in the plot 
deviate from the locus toward the upper left,
implying more concave spectral shapes than single power-laws,
their behavior can be explained away by contamination 
in M1-band by thermal signals (i.e., the 2nd parameter),
which causes a spectral steepening toward lower energies
(reduced M2/M1 ratios).
In fact, these deviating events exhibit L/M2 ratios of $14.4 \pm 8.5$,
which are much higher than those of the remaining ones, $2.9 \pm 2.2$
(here, the error represents the standard deviation among each subsample);
instead, the PC1/M2 ratios of these deviating events, $1.8 \pm 1.0$,
are comparable to those of the rest, $1.5 \pm 0.9$.

On the PC1-H-M2 color--color plane of figure~\ref{fig:color_color}c,
the 40 flares again exhibit two-dimensional scatter.
This implies 
that the gamma-ray to hard X-ray ratio (i.e., the 3rd parameter) 
remains undetermined,
even when the hard X-ray spectral slope (represented by H/M2) is specified.
In other words, the gamma-ray peak counts can deviate
significantly from a simple extension of the hard X-ray continuum,
and the deviation is apparently affected by some mechanism
other than that controlling the hard X-ray spectrum.
We hence regard the hard X-ray spectral slope as the 4th parameter,
which is independent of the 3rd one.

Finally, we show a scatter plot in panel (d)
among the highest three energy bands (HXT-H, GRS-PC1, and GRS-PC2).
Thus, the PC2/PC1 ratio stays relatively constant within a factor of $\sim 3$,
while the GRS-PC1/HXT-H ratio (the 3rd parameter) varies by an order of magnitude.
We hence regard the gamma-ray spectral slope as being approximately constant.
The gamma-ray photon indices implied by the PC2/PC2 ratios
distribute in the range 1.5--2.5 (see dashed lines in panel d),
which are generally flatter, and less scattered,   
than those found in the hard X-ray range ($\Gamma$ = 1.5--3.5; panel b).
This is consistent with the inference derived above from panel (c).
(Here we do not try to convert the GRS vs. HXT count ratios into 
a power-law index linking the hard X-ray and gamma-ray regions,
since the relative calibrations of the two instruments are subject to 
the uncertainty mentioned in subsection 2.2.)
However, the relative constancy of the 
gamma-ray slope can be subject to a selection bias,
because those flares of which the gamma-ray intensity is below the GRS threshold
could have rather different (possible much steeper) gamma-ray slopes.

\subsection{Selection Effects}

We have so far argued 
that the shape of higher energy SEDs of our sample flares may be 
characterized primarily by the two empirical parameters:
the gamma-ray to hard X-ray intensity ratio (the 3rd parameter),
and the hard X-ray spectral slope (the 4th parameter).
These results, however, could be artifacts introduced,
e.g., by the different hard X-ray and gamma-ray thresholds.
In order to examine this issue, 
we below consider several additional analyses.

First, we notice in figure~\ref{fig:scatter}d 
a very good proportionality between the GRS-PC1 and HXT-H counts
to hold over two orders of magnitude,
reflecting the 1st parameter,
together with a relatively homogeneous data scatter around it
(i.e., the 3rd parameter).
In fact, the best-fit correlation slope 
between the two quantities is 0.84,
and it becomes 0.98
if we exclude 5 events in the lower-left end of figure~\ref{fig:scatter}d.
This large-scale proportionality and the homogeneous scatter around it mean
that the hard X-ray vs. gamma-ray relation 
is rather independent of the flare size,
and hence is not significantly affected 
by the hard X-ray and gamma-ray thresholds to be employed.

As a next step, we examined what happens 
if we actually lower the gamma-ray threshold.
Specifically, we searched our Preliminary Sample (178 flares)
for those events which exhibit {\it weaker} gamma-ray peak signals,
with the background-subtracted GRS-(1+2) PC1 peak counts
in the range between 50 (typically 2$\sigma$) and 99 cts s$^{-1}$.
We found 15 such events, and call them collectively ``Additional Sample''.
In figure~\ref{fig:scatter},
we show a plot of these events with filled circles;
they simply form a smooth continuation from the Main Sample flares.
In the color--color diagrams (figures~\ref{fig:color_color}a through \ref{fig:color_color}c),
the two samples do not show any noticeable systematic differences, either.
Therefore, our results derived so far are robust
against a factor of 2 reduction in the GRS threshold.
We further searched the overall GRS data base for any other positive signals,
ignoring the HXT criterion,
but found none except those already registered in our Main and Additional Samples.
In short, any Yohkoh gamma-ray event is detected at the same time by the HXT;
the HXT is much more sensitive to solar flares than the GRS.

\begin{figure}[phbt]
\centerline{\includegraphics[width=14cm,angle=0,clip]
{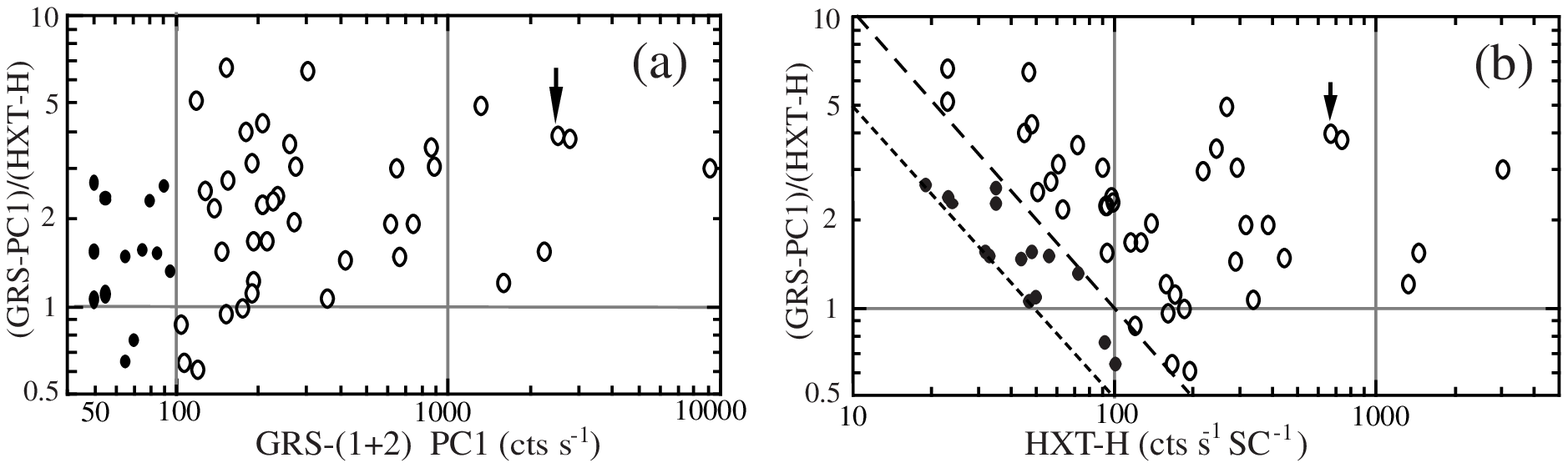}}
\caption{
(a) Scatter plot between the GRS-(1+2) PC1 peak counts
and the PC1/H count ratios.
The meanings of the open and filled symbols are 
the same as in figure~\ref{fig:scatter}.
The 1998 August 18 flare is specified again by an arrow.
(b) That between the HXT-H counts and the PC1/H count ratios.
The region to the left of the dashed lines is not covered by respective samples,
because of the threshold in the GRS-PC1 intensity.
}
\label{fig:bias}
\end{figure}

In order to examine more systematically 
how the GRS threshold affects the results obtained so far,
we show in figure~\ref{fig:bias}
two color--intensity relations of the 40 plus 15 flares.
As can be seen in panel (a),
the PC1/H ratio depends to some extent on the PC1 counts
($2.5 \pm 1.5$ for Main Sample and $1.6 \pm 0.6$ for Additional Sample),
but the dependence is much weaker than the overall scatter in PC1/H,
which persists across the full range of PC1 intensity down to our sensitivity limit.
Similarly, as shown in figure~\ref{fig:bias}b,
the overall distribution of the PC1/H ratio 
exhibits no particular dependence on the HXT-H counts,
except for the absence of events toward the lower left of the dotted lines;
this is due to those weak flares
that hit the HXT count threshold (100 c s$^{-1}$ SC$^{-1}$ in any band),
but not the GRS PC1 criterion (100 or 50 c s$^{-1}$).

\begin{figure}[phbt]
\centerline{\includegraphics[width=12cm,angle=0,clip]
{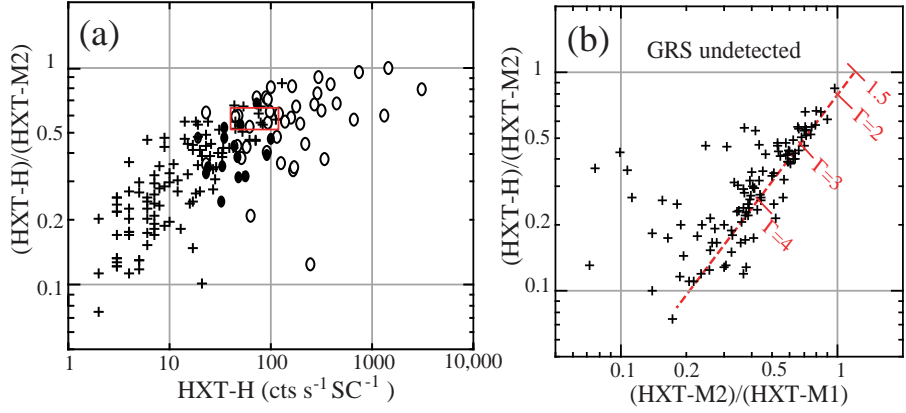}}
\caption{
(a) Hard X-ray spectral hardness (represented by the HXT-H M/M2 ratio)
of the 178 flares in our Preliminary Sample,
shown as a function of the HXT-H peak counts.
Among them, 
the 40 flares with significant GRS detections (i.e., Main Sample)
are indicated by open circles, 
and the 15 events with marginal GRS detections (i.e., Additional Sample)
with filled circles.
The crosses represent the remaining 123 flares without positive GRS signals.
(b) Same as figure~\ref{fig:color_color}b,
but plotted for the 123 GRS-undetected flares in Preliminary Sample.
In panel (a), they are indicated also by crosses.
}
\label{fig:color_color_nogamma}
\end{figure}

Then, how about the majority of flares below the GRS detection threshold?
We accordingly returned to our Preliminary Sample, 
consisting of the 178 flares (section~3),
which exceed the HXT threshold
but mostly lack positive gamma-ray signals.
As a representative color--intensity plot without invoking the GRS data,
we show in figure~\ref{fig:color_color_nogamma}a 
their H/M2 ratios as a function of the H-band peak counts.
Thus, the flares form a single, elongated, two-dimensional distribution,
with a clear positive correlation between the two quantities.
The correlation is not surprising,
since a flare with a harder non-thermal spectral slope
is naturally expected to produce higher HXT-H counts.
If the GRS were more sensitive,
the PC1/H vs. PC1 color--intensity diagram of 
figure~\ref{fig:bias}a would reveal a similar trend.

We also find in figure~\ref{fig:bias}a 
that the two GRS-detected samples (Main and Additional) 
occupy the upper right end of the distribution;
therefore, our Main Sample clearly selects those flares
with a larger size (the 1st parameter)
and a flatter hard X-ray slope (the 3rd parameter) for obvious reasons.
Nevertheless, we observe a significant mixing of the 123 gamma-ray 
undetected flares (crosses) with the gamma-ray detected ones (circles);
this would not occur if the gamma-ray intensity were
uniquely determined by these two parameters alone.
In other words, 
some flares have detectable gamma-ray signals, while others do not,
even though they have comparable hard X-ray intensities 
and similar hard X-ray slopes. 
Specifically, 
a representative region in figure~\ref{fig:color_color_nogamma}a
marked by a red rectangle
contains 6 gamma-ray detected and 6 non-detected flares.
Even though their H-band peak counts are all in a range of 45--115
and their H/M2 ratios are clustered in 0.48--0.63,
their PC1/H ratios scatter 
by more than an order of magnitude from 4.3 to $<0.35$;
this significantly exceeds 
what can be explained (up to a factor of $\sim 4$) 
by their differences in the hard X-ray data.
Thus, the inclusion of gamma-ray upper limits
strengthen our inference,
that the flare gamma-ray intensity can deviate 
significantly from the extrapolated hard X-ray continuum.

Figure~\ref{fig:color_color_nogamma}b shows the M2/M1 vs. H/M2 color--color diagram,
covering the 123 gamma-ray undetected flares in our Preliminary Sample. 
Compared with figure~\ref{fig:color_color}b,
which represents the gamma-ray detected events,
the overall data distribution is clearly shifted toward softer spectral slopes,
typically by one in $\Gamma$;
this is what is expected from figure~\ref{fig:color_color_nogamma}a.
Nevertheless, the plot recovers the essential features of figure~\ref{fig:color_color}b,
that a majority of events aligns one-dimensionally along the single power-law locus,
with a minority spreading toward lower M2/M1 regions
again due to thermal contamination.
Incidentally, about 10\% of the events are deviated 
to the lower right of the locus, 
implying convex spectral shapes.
Excluding these two types of exceptional cases, 
the nearly power-law shaped hard X-ray continua (i.e., the 3rd parameter)
are widely observed among our sample flares,
without regard to the presence/absence of positive GRS signals.

\begin{figure}[phbt]
\centerline{\includegraphics[width=14cm,angle=0,clip]
{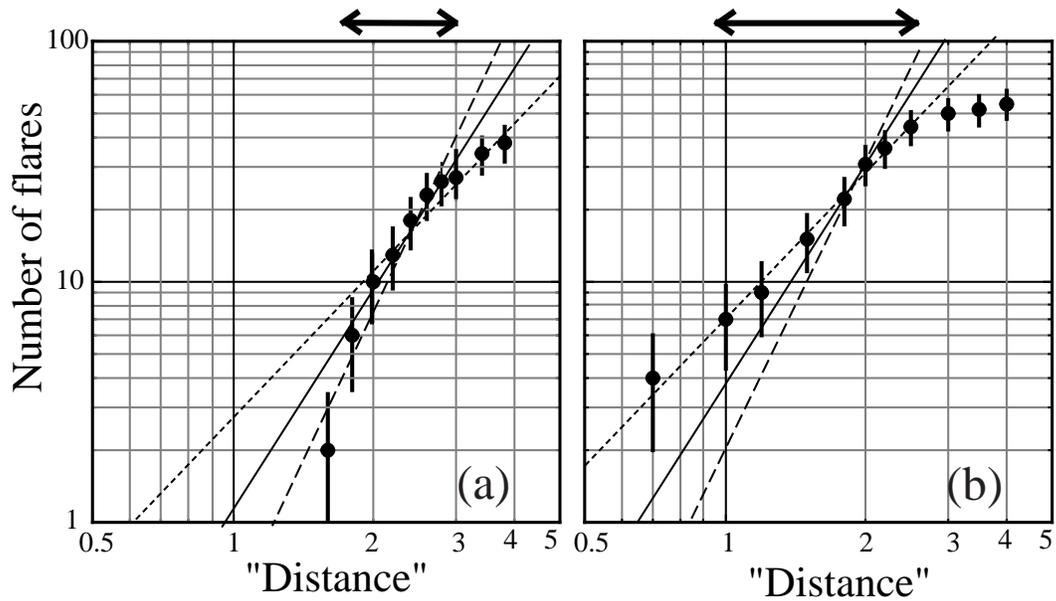}}
\caption{Results of a fractal dimension analysis.
The cumulative number of flares $C(<r)$,
of which the ``distance'' from the mean flare position
is smaller than a specified value $r$, 
is plotted as a function of $r$.
To guide the eyes, 
three straight lines with logarithmic slopes of 
2.0 (dotted), 3.0 (solid), and 4.0 (dashed) are drawn.
The arrow at the top of each panel 
shows the range used to define the slope.
(a) Calculated in the 7 dimensional space (GOES, HXT, GRS-PC1, and GRS-PC2)
over the 40 gamma-ray flares in Main Sample (see text subsection 4.2).
(b) Calculated in the 5 dimensional space (GOES and HXT) over 57 flares,
which exhibit significant HXT counts, but no GRS signals (see text subsection 4.4).
}
\label{fig:fractal}
\end{figure}

Finally, we performed a fractal dimension analysis to the gamma-ray undetected flares,
but further imposing a condition
that the event should have  $> 10$ counts in all four of the HXT bands,
so as to minimize any artifacts caused by Poisson errors.
This has reduced the number of flares from 123 to 57.
Figure~\ref{fig:fractal} shows a curve $C(>r)$ of these 57 flares,
calculated in a 5-dimensional embedding space
spanned by the GOES class and the four HXT bands.
Thus, $C(>r)$  is clearly flatter than that in figure~\ref{fig:fractal}a,
and the power-law fit over the range of $0.9<r< 2.7$ 
gives a slope of $1.9 \pm 0.4$,
which is smaller by about 1 than that in panel (a).
This is quite reasonable,
because the sample now lacks one particular degree of freedom,
namely the gamma-ray to hard X-ray intensity ratio (the 3rd parameter).

From these examinations,
we conclude that the selection bias
does not affect the two principal results of our analysis;
the prevalence of the power-law shaped hard X-ray SED,
and the significant scatter of the gamma-ray intensity 
relative to the extrapolation of the hard X-ray continuum.

\subsection{Wide-Band Spectra}
Although a detailed analysis of the individual flares is 
beyond the scope of the present paper,
looking at several representative spectra would help
us to understand the statistical results obtained so far.
Accordingly, in figure~\ref{fig:wide_spec}
we show several examples of the GRS pulse-height data
on which the HXT four-band data are approximately aligned.
The GRS spectra were obtained 
by integrating the GRS PH-L data over the period 
when the GRS-1 signal exceeds the background by more than 3$\sigma$,
and then subtracting the background,
utilizing pre- and post-flare periods.
Similarly, we accumulated the HXT counts 
over the same period as the GRS integration,
subtracted the post-flare background, 
summed the results over the 64 subcollimators,
and finally corrected the sum in each band
for the bandwidth and a typical detection efficiency there;
the latter was calculated using the HXT response 
to a power-law spectrum of photon index 3.0.
As a result, the arrangement of the four-band HXT data points 
roughly represents the power-law slope of the incident hard X-rays.

A few remarks should be added to figure~\ref{fig:wide_spec}.
Firstly, the relative flux alignments 
between the GRS and HXT are approximately correct,
except for the factor of $\sim 2$ uncertainty 
in their mutual calibration (subsection 2.2).
Secondly, the SEDs displayed in this figure are slightly different 
from those implied by the counts in table~\ref{tbl:sample},
because the former refers to time averages,
while the latter to the peak values in the individual energy bands.
Finally, in several cases we observe nuclear lines,
including the 2.23 MeV neutron-capture lines
and several features attributable to the nuclear de-excitation process.
This is indeed another important subject,
but is beyond the scope of the present paper.

Figure~\ref{fig:wide_spec}a compares the 1998 August 18 limb flare
with the 2000 July 14 event that occurred near the solar disk center.
Their gamma-ray to hard X-ray flux ratios are significantly different,
although their SEDs (except the normalization) are 
close to each other over the 25--100 keV range
and their gamma-ray spectral slopes are rather similar.
Panel (b) provides a comparison between two disk flares,
having the flattest and the steepest hard X-ray slopes 
in terms of the H/M2 ratio among our Main Sample.
Nevertheless, they are quite similar in  
the gamma-ray to hard X-ray ($\sim 100$ keV) flux ratio,
as well as in the gamma-ray spectral slope.
These flares thus exemplify the mutual independence 
of the 3rd and 4th parameters.

\section{Discussion}

Through a fractal dimension analysis of 
the 40 gamma-ray emitting solar flares (Main Sample),
we arrived at an inference 
that their wide-band SEDs have 
at least four degrees of freedom.
Furthermore, the color--color plots allowed us to
assign the following meanings to the four parameters:
\begin{enumerate}
\item The overall flare size.
\item The dominance of the thermal signal over the non-thermal one.
\item The gamma-ray vs. hard X-ray relative intensities.
\item The spectral slope of non-thermal hard X-rays.
\end{enumerate}
Precisely speaking, 
there must be an additional degree of freedom,
i.e., the temperature of the thermal emission,
but our energy bands are probably too coarse
for this effect to be noticed.
Furthermore, the gamma-ray spectral slope 
could be yet another independent parameter, 
although our sample does not clearly show its presence.

While the 1st, 2nd, and 4th parameters represent well-recognized flare properties,
the reality of the 3rd parameter (the gamma-ray vs. hard X-ray flux ratio),
and its independence from the 4th parameter, are both non-trivial results.
Accordingly, in subsection~4.4
we carefully examined these results for selection effects,
incorporating weaker gamma-ray flares
and those with gamma-ray upper limits.
As a result, we confirmed
that our conclusion derived from Main Sample can be extended 
to cover a large fraction of the entire hard X-ray flare sample.
In other words, the selection bias is small,
and does not affect our conclusion.

\begin{figure}[phbt]
\centerline{\includegraphics[width=14cm,angle=0,clip]
{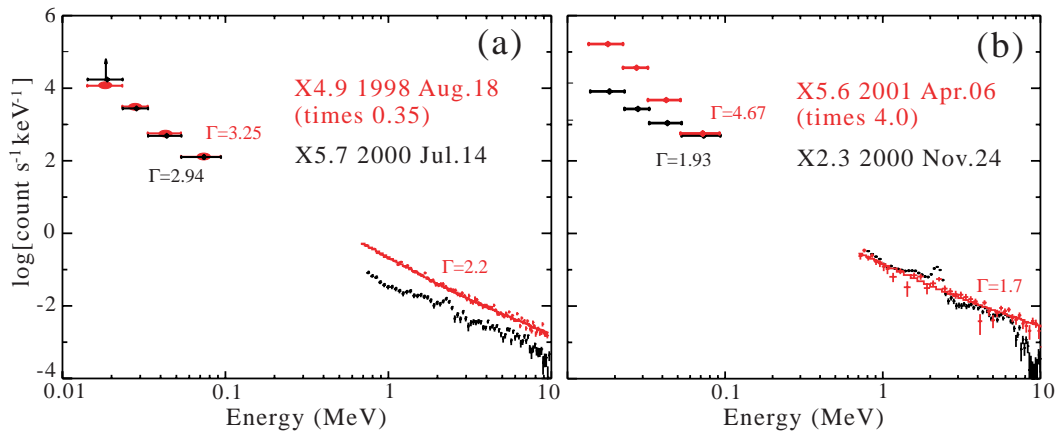}}
\caption{Examples of the background-subtracted 
wide-band spectra of our sample flares,
obtained as described in the text.
The GRS pulse-height data (0.7--10 MeV) are presented 
without any correction for the instrumental response.
When the spectra are featureless, 
the prediction of the best-fit power-law model is also shown,
with the photon index given in the figure.
Four horizontal bars show the background-subtracted four-band HXT data,
obtained as described in the text.
The photon index, $\Gamma$, implied by the H-band to M2-band ratio,
is given in the figure.
(a) A comparison of the 1998 August 18 limb flare (in red),
with the 2000 July 14 disk flare (in black).
Counts of the former are multiplied by 0.35.
The L-band data of the latter flare suffered from telemetry overflows.
(b) The same as panel (b), 
but for the 2001 April 6 flare (red)
and the 2000 November 24 flare (black).
Counts of the former are multiplied by 4.0.
}
\label{fig:wide_spec}
\end{figure}

\begin{figure}[phbt]
\centerline{\includegraphics[width=8cm,angle=0,clip]
{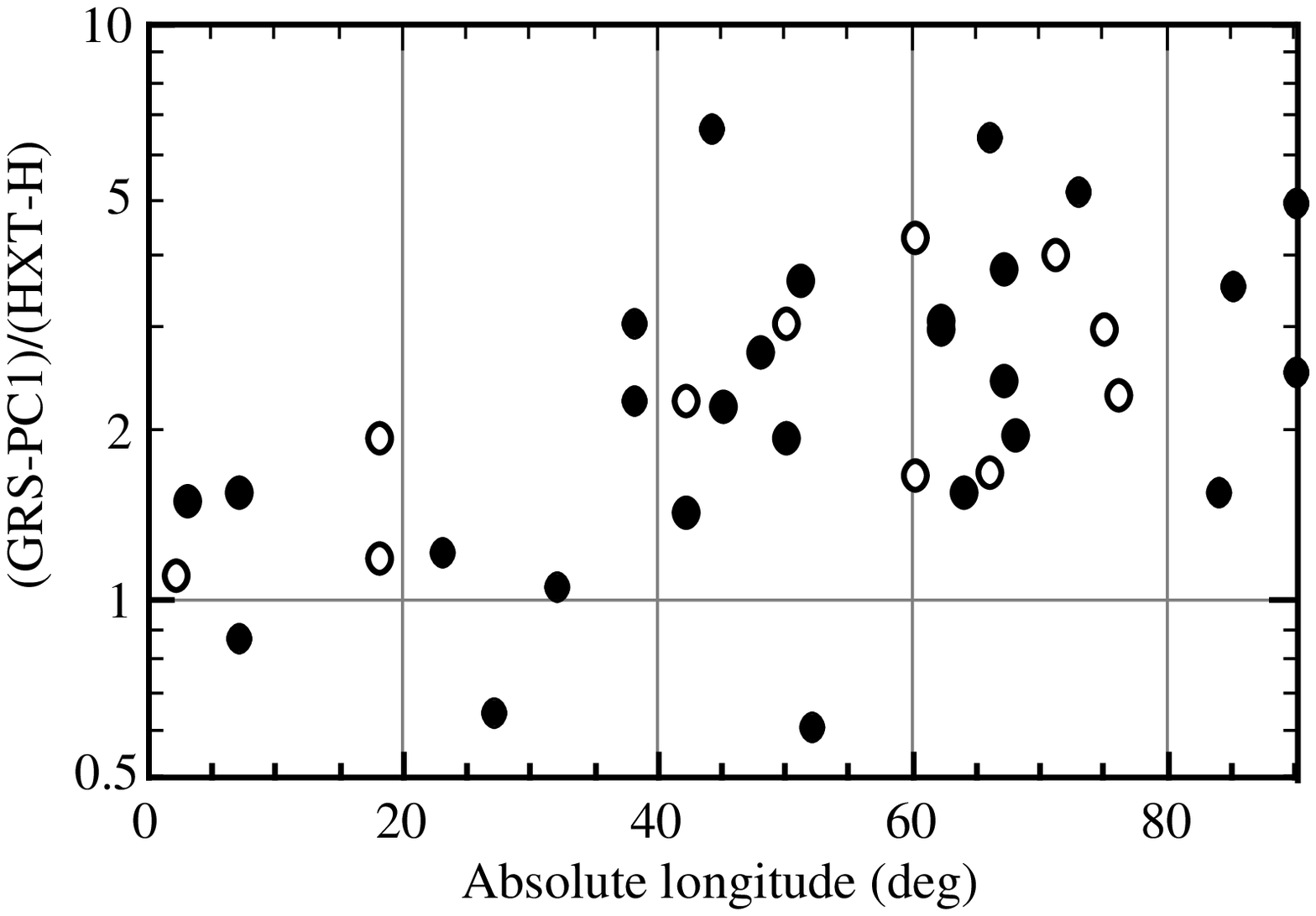}}
\caption{PC1/H ratio of the Main Sample flares, 
plotted as a function of their absolute solar longitudes.
The open and filled circles represent the flares 
recorded in the solar cycles 21 and 22, respectively.
}
\label{fig:longitude}
\end{figure}

It is known that limb flares show systematically 
stronger gamma-ray emission than disk flares 
(e.g., McTiernan, Petrosian 1991; Vestrand, Forrest 1992), 
presumably due to the energy-dependent anisotropy.
Since the electron bremsstrahlung emissivity is relativistically forward-peaked,
the gamma-ray photons of a disk flare would 
have to be Compton back-scattered to reach us,
and hence lose a larger fractional energy 
than those of limb flares (Kotoku 2004).
Then, the scatter in the PC1/H ratio 
in our sample may simply reflect this effect.
In fact, in figure~\ref{fig:wide_spec}, 
the limb flare of 1998 August 18 exhibits 
stronger gamma-rays (relative to the hard X-rays)
than the 2000 July 14 flare that occurred near the disk center.
In order to examine this possibility,
we plot in figure~\ref{fig:longitude} the PC1/H ratios of the 40 flares 
against the absolute solar longitudes of their occurrence.
There, we separately treated the flares detected before 1992 and those after 1997,
for fear that the GRS gain change (subsection~2.2) is 
artificially enhancing the scatter in the PC1/H ratio.
Figure~\ref{fig:longitude} reconfirms the trend of increasing gamma-ray intensity
(relative to the hard X-ray signal) toward the solar limbs,
but the effect, at most a factor of $\sim 2.5$,
is too small to explain away the 3rd parameter.
For example, the difference between the two flares shown in figure~\ref{fig:wide_spec}a
could partially be due to the longitudinal effects,
but their difference in the gamma-ray to hard X-ray flux ratio 
(by a factor of $\sim 8$)
may be too large to be explained in that way.
We also confirm that the GRS gain change has a negligible effect, 
as we do not observe significant differences between 
the two sabsamples corresponding to the solar cycles 21 and 22.

The observed scatter in the PC1/H ratio 
suggests some intrinsic physics involved in the flare dynamics.
One simple possibility is
that the maximum energies of electrons may vary from flare to flare,
causing the gamma-ray flux to deviate from the hard X-ray extension.
While this may well be taking place,
it alone cannot explain our results, 
since the measured gamma-ray flux in many of our Main Sample flares 
considerably exceeds the power-law extrapolation from the hard X-ray spectrum,
and the gamma-ray slope is often significantly flatter than the hard X-ray slope
(see figure~\ref{fig:wide_spec}).
Some fraction, if not all, of the excess gamma-ray signals 
may be attributed to nuclear lines, 
such as can be observed in figure~\ref{fig:wide_spec}b;
this raises an interesting possibility
that the 3rd parameter might be related to the relative dominance
between the accelerations of electrons and ions.
Even considering the electron bremsstrahlung alone,
there is a possibility
that flares generally involve multiple populations of 
energetic electrons with different spectra;
for example, the 2001 April 06 flare 
(figure~\ref{fig:wide_spec}b; without noticeable nuclear lines)
might involve a soft electron population emitting the steep-spectrum hard X-rays,
and a harder electron population responsible for the flatter-spectrum gamma-ray emission.
Then, the 3rd parameter may be interpreted as representing
the relative contributions from these different electron populations.
As yet another possibility,
a single electron population may produce different photon spectra,
depending on the physical condition (thin-target of thick-target)
of the emission region,
as well as the path of the photon propagation;
the observed longitudinal effect is one such example.
Therefore, the 3rd parameter could be a composite quantity
consisting of multiple factors,
of which the effects are difficult to single out individually.
We defer further examination of the 3rd parameter to our later publications.

\clearpage
\section*{References}

\re Acton, L., et al. 1992, Science, 258, 618
\re Carmichael, H. 1962, Space Sci. Rev., 1, 28
\re Hirayama, T. 1974, Solar Phys., 34, 323
\re Inda-Koide, M. 1994, PhD Thesis, the University of Tokyo
\re Inda-Koide, M., Makishima, K., Kosugi, T., \& Kaneda, H. 1995, PASJ, 47, 661
\re Kopp, R. A., \& Pneuman, G. W. 1976, Solar Phys., 50, 85
\re Kosugi, T., et al. 1991, Solar Phys., 136, 17
\re Kosugi, T., et al. 1992, PASJ, 44, L45
\re Kotoku, J. 2004, PhD Thesis, the University of Tokyo
\re Masuda, S., 1994, PhD Thesis, the University of Tokyo
\re Masuda, S., Kosugi, T., Hara, H., Sakao, T., Shibata, K., 
    \& Tsuneta, S. 1995, PASJ, 47, 677
\re Masuda, S., Kosugi, T., Hara, H., Tsuneta, S., \&  Ogawara, Y. 1994, 
    Naturre 371, 495
\re Matsumoto, Y. 2002, PhD Thesis, the University of Tokyo
\re McTiernan, J. M., \& Petrosian, V. 1991, ApJ, 379, 381
\re Ogawara, Y., et al. 1992, PASJ, 44, L41O
\re Petrosian, V., Donaghy, T. Q., \& McTiernan, J. M. 2002, ApJ, 569, 459
\re Sakao, T. 1994, PhD Thesis, the University of Tokyo
\re Sakao, T., et al. 1992, PASJ, 44, L83
\re Sato, J. 1997,  PhD Thesis, the Graduate University for Advanced Studies
\re Sato, J.,  Kosugi, T., \& Makishima, K. 1999, PASJ, 51, 127
\re Sturrock, P. A., \& Coppi, B. 1966, ApJ, 143, 3
\re Vestrand, W. T., \& Forrest, D. J. 1992,  BAAS, 24, 802
\re Yoshimori, M., et al. 1991, PASJ, 44, L51

\clearpage
\vspace*{5cm}
\begin{table}[hbt]
\begin{small}
\begin{center}
\caption{Energy ranges (in MeV) of the GRS data, applicable for the solar cycle 22.}
\label{tbl:GRS}
\begin{tabular}{lccc}
\hline\hline
Data       & $\Delta T~^{*}$ & GRS-1  &  GRS-2 \\ 
\hline 
PC1		& 0.25	 & 0.22--1.43	& 0.49--2.15\\
PC2		& 0.25		& 1.43--6.21	& 2.15--7.65	\\
PC3		& 0.5			& 6.21--8.87	& 7.65--11.8	 \\
PC4		& 0.5			& 8.87--17.1	& 11.8--23.0	 \\
PC5		& 0.5		 & 10--30		    & 8--30	 \\
PC6		& 0.5			& 30--100	   & 30--100		 \\
\hline
\multicolumn{3}{l}{$^{*}$  Time resolution in seconds.}\\
\end{tabular}
\end{center}
\end{small}
\end{table}

\begin{table}[p]
\begin{center}
{\small 
\caption{The 40 flares in Main Sample, with significant hard X-ray and gamma-ray emissions.}
\label{tbl:sample}
\begin{tabular}{llrrrrrrrrl}
\hline\hline
GOES$^*$ & yy-mm-dd	& \multicolumn{4}{c}{HXT $^{\dagger}$}
                          	& \multicolumn{3}{c}{GRS-(1+2) $^\ddagger$}	& Position\\
                   \cline{3-6}          \cline{7-9}
     &           &  L    & M1   & M2   & H    & PC1   & PC2  & PC3 \\  
\hline
X14  & 01-apr-15	& 24123 & 5917 & 1981 & 246	&  867  &  201 & 9   & S20W85 \\
X9.4 & 97-nov-06	& 19765 & 5569 & 3831 & 3069	& 9139  & 1449 & 53  & S18W62 \\
X6.1 & 91-oct-27	& 3802  & 2586 & 2193 & 1323	& 1586  &  266 & 42  & S12E18 \\
X5.7 & 00-jul-14 	&$>$4096& 870  & 645  &  444	&  662  & 128  & 14  & N13W03 \\
X5.6 & 01-apr-06	& 13361 & 2829 & 886  &  337	&  358  & 66   & 10  & S13E32 \\
X4.9 & 98-aug-18	& 10346 & 2352 & 1144 &  661	& 2576  & 1574 & 307 & N30E84 \\
X4.0 & 00-nov-26	& 6081  & 1052 & 224  &   92	&  206  & 62   & 14  & N18W38 \\
X3.7 & 98-nov-22	& 6234  & 1213 & 393  &  268	& 1320  & 324  & 22  & S31W90 \\
X3.3 & 98-nov-28	& 2593  & 946  & 302  &   63	&  183  & 44   & 16  & N20E45 \\
X2.7 & 98-may-06	& 2911  & 448  & 108  &   47	&  302  & 50   & 10  & S15W66 \\
X2.6 & 97-nov-27	& 4354  & 720  & 168  &   94	&  146  & 46   & 8   & N18E64 \\
X2.3 & 00-nov-24	& 2730  & 1256 & 1444 & 1453	& 2252  & 376  & 10  & N22W07 \\
X2.2 & 91-dec-03	& 4086  & 1286 & 488  &  217	&  644  & 140  & 8   & N18E75 \\
X1.9 & 00-nov-25	& 1958  & 370  & 268  &  158	&  192  & 60   & 7   & N20W23 \\
X1.9 & 00-jul-12	& 3587  & 963  & 491  &  165	&  106  & 39   & 9   & N17E27 \\
X1.5 & 91-nov-15	& 1254  & 701  & 499  &  318	&  610  & 108  & 12  & S14W18 \\
X1.2 & 00-sep-30	& 1629  & 483  & 133  &   51	&  128  & 32   & 6   & N07W90 \\
X1.1 & 00-mar-02	& 1350  & 515  & 351  &  195	&  119  & 45   & 6   & S14W52 \\
X1.0 & 92-jan-26	& 743   & 614  & 350  &  127	&  214  & 42   & 8   & S16W66 \\
X1.0 & 99-aug-02	& 1749  & 360  & 132  &   57	&  155  & 45   & 7   & S22W48 \\
M9.8 & 99-aug-20	& 1759  & 706  & 773  &  742	& 2806  & 504  & 68  & S26E67 \\
M9.0 & 91-nov-02	& 585   & 377  & 239  &  115	&  192  & 37   & 6   & S12W60 \\
M8.2 & 00-nov-25	& 163   & 106  & 98   &   72	&  260  & 66   & 12  & N09E51 \\
M8.0 & 98-dec-18	& 482   & 352  & 243  &  138	&  270  & 46   & 8   & N21E68 \\
M8.0 & 00-jul-25	& 571   & 262  & 195  &  120	&  104  & 46   & 8   & S01W07 \\
M7.9 & 91-nov-10	& 366   & 155  & 122  &   90	&  274  & 72   & 8   & S14E50 \\
M7.0 & 92-feb-14	& 1061  & 893  & 485  &  170	&  190  & 69   & 5   & S12E02 \\
M6.8 & 92-jul-16	& 191   & 115  & 79   &   48	&  206  & 58   & 6   & S10W60 \\
M6.7 & 01-mar-10	& 512   & 418  & 377  &  290	&  416  & 94   & 8   & N27W42 \\
M4.5 & 99-dec-28	& 307   & 456  & 457  &  386	&  744  & 210  & 8   & N25W50 \\
M3.8 & 00-mar-03	& 241   & 174  & 113  &   61	&  189  & 61   & 9   & S13W62 \\
M3.2 & 92-sep-10	& 122   & 139  & 129  &   93	&  208  & 42   & 6   & N18E42 \\
M3.1 & 98-aug-14	& 148   & 60   & 37   &   23	&  118  & 50   & 6   & S25W73 \\
M2.5 & 01-jun-05	& 126   & 120  & 64   &   23	&  152  & 44   & 8   & S18E44 \\
M2.4 & 00-jul-17	& 209   & 307  & 321  &  292	&  888  & 126  & 4   & S17E38 \\
M1.6 & 01-aug-31 	& 140   & 200  & 195  &  162 &  150  & 65   & 8   & N15E37 \\
M1.6 & 01-nov-12 	& 322   & 341  & 269  &  179 &  175  & 85   & 8   &        \\
M1.5 & 99-dec-18	& 339   & 239  & 153  &   97	&  234  & 74   & 12  & N20E67 \\
M1.4 & 91-nov-09	& 85    & 101  & 75   &   45	&  180  & 70   & 14  & S16W71 \\
M1.4 & 91-dec-15	& 142   & 129  & 121  &   99	&  227  & 33   & 7   & S12E76 \\
\hline
\end{tabular}
}
\end{center}
\begin{description}
\vspace{-3mm}
{\footnotesize
\item[$^{*}$]
The GOES class, defined in the 1.5--12 keV band.
\vspace{-2mm}
\item[$^{\dagger}$] The hard X-ray peak counts in the four HXT energy bands,
after background subtraction, in units of c s$^{-1}$ SC$^{-1}$.
\vspace{-2mm}
\item[$^\ddagger$] The gamma-ray peak counts in the four GRS-1 plus GRS-2 PC bands,
after background subtraction, in units of c s$^{-1}$.
}
\end{description}
\end{table}

\clearpage
\vspace*{5cm}
\begin{table}[htbp]
\begin{center}
\caption{Correlation coefficients among peak counts in seven energy bands,
for the 40 gamma-ray flares of table~\ref{tbl:sample}. 
}
\vspace{3mm}
\label{tbl:correlation_coef}
\begin{tabular}{lccccccc}
\hline\hline
           & \multicolumn{4}{c}{HXT} & &\multicolumn{2}{c}{GRS-(1+2)}\\
           \cline{2-5}\cline{7-8}
	          &  L  &  M1	  &  M2	  &  H    & &PC1	 &  PC2  \\
\hline 
GOES	      &0.920& 0.873	& 0.721	& 0.548	& &0.478	& 0.470 \\
HXT-L	     & ---	& 0.925	& 0.737	& 0.553	& &0.480	& 0.490 \\
HXT-M1	    & ---	& ---	  & 0.911	& 0.737	& &0.623	& 0.603 \\
HXT-M2	    & ---	& ---	  & ---	  & 0.926	& &0.771	& 0.732 \\
HXT-H	     & ---	& ---	  & ---	  & ---	  & &0.854	& 0.806 \\
GRS-PC1    & ---	& ---	  & ---	  & ---	  & &---	  & 0.948 \\
\hline
\end{tabular}
\end{center}
\end{table}

\clearpage
\vspace*{5cm}

\clearpage
\vspace*{5cm}

\clearpage
\vspace*{5cm}

\clearpage
\vspace*{5cm}

\clearpage
\vspace*{5cm}

\clearpage
\vspace*{5cm}

\clearpage
\vspace*{5cm}

\end{document}